\documentclass[aps,prl,twocolumn,groupedaddress,floatfix]{revtex4}
\usepackage{graphicx}
\usepackage{amsmath}
\usepackage{hyperref}

\begin{document}

\newcommand{\barium} {$^{138}$Ba$^{+}$}
\newcommand{\baplus} {Ba$^{+}$}
\newcommand{\cytt} {Cyt$^{12+}$}
\newcommand{\cyts} {Cyt$^{17+}$}
\newcommand{\cytn} {Cyt$^{n+}$}

\title{Translational cooling and storage of protonated proteins in an
ion trap at subkelvin temperatures}
\author{D. Offenberg, C. B. Zhang, Ch. Wellers, B. Roth, and S. Schiller}
\affiliation{Institut f\"{u}r Experimentalphysik,
Heinrich--Heine--Universit\"{a}t D\"{u}sseldorf, 40225
D\"{u}sseldorf, Germany}


\begin{abstract}
Gas-phase multiply charged proteins have been sympathetically
cooled to translational temperatures below 1 K by Coulomb
interaction with laser-cooled barium ions in a linear ion trap. In
one case, an ensemble of 53 cytochrome c molecules (mass $\simeq$
12390 amu, charge +17 $e$) was cooled by $\simeq 160$ laser-cooled
barium ions to less than 0.75 K. Storage times of more than 20
minutes have been observed and could easily be extended to more
than an hour. The technique is applicable to a wide variety of
complex molecules.
\end{abstract}

\maketitle

In the last years, the development of techniques for producing
cold molecules ($<1$~K) has moved into focus. Remarkable advances
in this young field of quantum physics are enabling studies of
light-molecule, atom-molecule and molecule-molecule interactions
in a temperature regime not accessible previously. Cold molecule
types produced so far range from neutral diatomic molecules
produced by photoassociation in ultracold atomic gases
\cite{Fioretti1998} to polar few-atom molecules slowed down by
Stark deceleration \cite{Cromp2001} and charged diatomic to
polyatomic molecules sympathetically cooled by laser-cooled atomic
ions in ion traps \cite{Drewsen2004,Roth2005,Ostendorf2006}. The
heaviest cold molecular species prepared had a mass of 410~amu.

It is of interest to develop methods that allow a similar control
over much heavier particles like anorganic clusters or
biomolecules, consisting of hundreds or thousands of atoms. So
far, such complex systems have been studied in ion traps using
buffer gas cooling \cite{Gerlich1992,Asvany2005}. The buffer gas
thermalizes with a cryogenically cooled chamber containing the
trap and collisionally cools the trapped ions internally and
translationally to temperatures of $\sim10$~K \cite{Stearns2007a}.
Recently we have reported on a technique for cooling polyatomic
molecular ions that allows one to achieve significantly lower
translational temperatures and long storage times of up to hours.
As examples, singly protonated dye (mass 410 amu)
\cite{Ostendorf2006} and glycyrrethinic acid molecules (mass 471
amu) \cite{Offenberg2008} produced by electrospray ionization
(ESI) have been sympathetically cooled to less than 150 mK by
their mutual Coulomb interaction with laser-cooled \barium~ions in
a linear quadrupole trap.

Here we demonstrate a major extension of this technique to much
heavier molecules (such as proteins), taking full advantage of the
ability of the ESI ion source to produce multiply charged ions
with mass-to-charge ratios suitable for conventional quadrupole
mass filters and linear ion traps \cite{Fenn1989,Schiller2003}. We
have produced, trapped, sympathetically cooled, and maintained
cold for extended intervals 12- and 17-fold protonated protein
molecules (cytochrome c, a protein consisting of about 100 amino
acids with a total mass of $\simeq 12390$~amu) using Doppler
laser-cooled \barium~ions in a linear quadrupole trap. The two
trapped species were detected and distinguished via excitation of
their specific motional resonances. The sympathetic cooling was
qualitatively proven by measuring the energy distribution of the
ions and quantitatively analyzed using molecular dynamics
simulations.

\begin{figure}[b]
    \includegraphics[width=7.25cm]{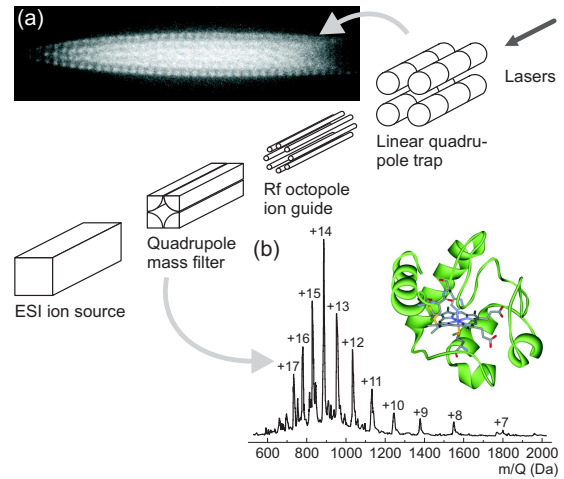}
    \caption{\label{fig:figure1} (Color online) Schematic overview of
    our setup. Singly or multiply protonated molecular ions from an ESI
    ion source are selected by a quadrupole mass filter and transferred
    via a RF octopole ion guide to a linear quadrupole trap in an UHV
    chamber. (a) CCD image of a typical trapped barium ion Coulomb
    crystal used as coolant for the molecular ions. (b) Secondary
    structure and ESI mass spectrum of cytochrome c showing the
    distribution of the different grades of protonation.}
\end{figure}

Our apparatus \cite{Ostendorf2006} is shown schematically in
Fig.~\ref{fig:figure1}. It consists of an ESI ion source for the
production of singly and multiply protonated molecular ions, a
quadrupole mass filter for the selection of specific molecular
species, a rf octopole ion guide to transfer the selected
molecular ions from the medium vacuum region of the apparatus to
an ultrahigh vacuum (UHV) chamber, and a linear quadrupole trap in
this UHV chamber to store the molecular ions for cooling and
further investigations. The preparation of the laser-cooled
\barium~ion ensembles used for sympathetic cooling of the
molecular ions as well as the required lasers (493~and 650~nm) is
described in \cite{Roth2005}. With sufficient cooling power the
barium ions arrange in ordered structures, so-called Coulomb
crystals, that can be imaged with a charge-coupled-device (CCD)
camera [Fig.~\ref{fig:figure1}~(a)]. The images provide
information not only on the fluorescing \barium~ions, but
indirectly also on the numbers and temperatures of the invisible
sympathetically cooled molecular ions. Molecular dynamics (MD)
simulations are used to derive these data from structural
deformations and sharpness of the barium ion Coulomb crystals
\cite{Zhang2007}.

Multiply protonated molecules of cytochrome c from horse heart
(Fluka BioChemika), (cyt c + $n$H)$^{n+}$ with grades of
protonation, $n=7-17$, have been produced by ESI from a solution
concentration of $10^{-5}M$ in 1:1 methanol:water. As the maximum
$n$ depends on the $p$H value of the solution, a solution with a
higher concentration of acetic acid of 2\% was used to produce
17-fold protonated cytochrome c molecules, here denoted as
Cyt$^{17+}$, and a lower concentration of 0.5\% was used to obtain
12-fold protonated molecules Cyt$^{12+}$.

The preparation procedures for sympathetically cooled Cyt$^{12+}$
and Cyt$^{17+}$ ion ensembles are identical. First, helium buffer
gas at room temperature is injected into the trap chamber to
pressures of $1-10\times10^{-5}$~mbar in order to subsequently
reduce the kinetic energy of the molecular ions from the octopole
ion guide so far that they can be trapped. After loading ions of
the desired molecular species for $1-10$~s depending on the
desired ion number and the ion flux (typically several thousand
ions per second reach the trap), the buffer gas is pumped away and
the background pressure in the trap chamber returns to
$<1\times10^{-9}$~mbar within a few minutes. Then, barium ions
from an evaporator oven are loaded into the trap and cooled by the
two cooling laser beams propagating along the trap axis. Impurity
ions generated during loading (such as molecular fragments or
CO$_{2}^{+}$ from the evaporator) can be removed from the trap by
applying an additional appropriately strong ac voltage to the trap
electrodes in order to excite the ions' specific motional
resonances beyond stability \cite{footnote1}. At a typical rf
amplitude of 450 V the radial resonance frequency of \barium~ions
is 118 kHz, while that of cytochrome ions is in the range of
10$-$22 kHz depending on their protonation grade. At the end of
the preparation procedure, two frequency scans (125$-$1000 and
110$-$40 kHz) are applied to purify the barium-cytochrome ion
ensemble.

\begin{figure}[t]
    \includegraphics[width=7.25cm]{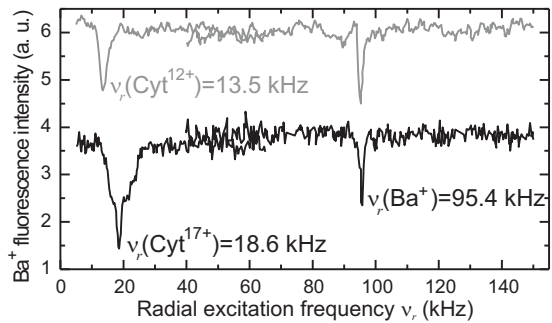}
    \caption{\label{fig:figure2} Radial excitation
    spectra of barium crystals containing differently protonated
    cytochrome ions. The upper (\cytt) and lower (\cyts) curves
    were acquired at the same trap parameters and yield the same
    barium frequency of 95.4~kHz used for calibration. The obtained
    cytochrome frequencies are well confirmed by their theoretical
    values.}
\end{figure}

The same principle of radial excitation can be used in a
nondestructive way to identify trapped ion species. For this
purpose, frequency scans with $\sim10$~times lower amplitudes than
for purification are performed while acquiring the barium ions'
fluorescence intensity with a photomultiplier. Under resonance
excitation of a particular ion species, its motion heats all
trapped ions due to their mutual Coulomb interaction. As a
consequence, the barium ions' fluorescence drops, leading to dips
in the radial excitation spectrum at the species-specific
resonance frequencies \cite{Baba2002,Roth2007}. Cyt$^{12+}$ and
Cyt$^{17+}$ in barium-ion Coulomb crystals have been identified
and distinguished via their radial resonance frequencies as shown
in Fig.~\ref{fig:figure2}. The upper and lower curves are composed
of two different scans: one from 5 to 65~kHz for the detection of
the molecular ions and one from 40 to 150~kHz with a 10 times
lower amplitude for the barium ions (in order to avoid crystal
destruction). As the rf amplitude was the same for all four scans
(362 V), the barium resonances coincide at
$\nu_{r}(\textrm{Ba$^{+}$})=95.4$~kHz and can be used for
calibration. With the resonance frequency $\nu_{r}\sim Q/m$ (for
$-a\ll q$ \cite{footnote1}) the expected frequencies for
(noninteracting) cytochrome ions can be calculated as
$\nu_{r}(\textrm{Cyt$^{n+}$})=\nu_{r}(\textrm{Ba$^{+}$})\cdot[Q(\textrm{Cyt$^{n+}$})/
Q(\textrm{Ba$^{+}$})]\cdot[m(\textrm{Ba$^{+}$})/m(\textrm{Cyt$^{n+}$})]$,
which yields 12.8~kHz for Cyt$^{12+}$ and 18.1~kHz for
Cyt$^{17+}$. These values are well confirmed by the experimental
results of 13.5 and 18.6~kHz. The small differences of less than
5\% can be ascribed to various factors like the direction and
amplitude of the frequency scans \cite{Baba2002} or motional
coupling effects \cite{Roth2007}.

\begin{figure}[t]
    \includegraphics[width=7.25cm]{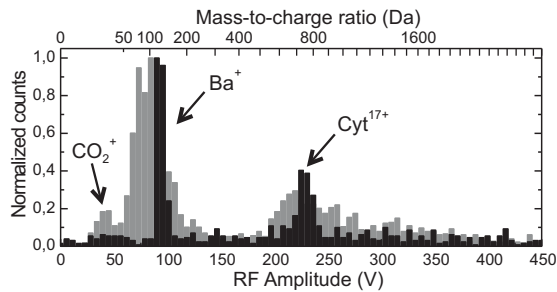}
    \caption{\label{fig:figure3} Ion extraction mass spectra of
    \baplus/\cyts ensembles at different temperatures. In the
    noncooled case (gray) the mass spectrum is broadened compared
    to the laser-cooled case (black). Here, not only the peak of the
    laser-cooled \baplus~ions, but also that of the simultaneously
    trapped \cyts~ions is narrower, which is a proof of their
    sympathetic cooling.}
\end{figure}

Additionally, we have implemented a destructive mass
identification technique which is based on a mass-selective
extraction of the ions from the trap. When decreasing the trap rf
amplitude $U_{\textrm{rf}}$ the trapping stability parameter $q$
\cite{footnote1} is reduced and ions escape the trap at a
mass-to-charge ratio dependent amplitude
$U^{\textrm{ex}}_{\textrm{rf}}\propto \sqrt{m/Q}$
\cite{footnote2}. In our setup an ion detector below the trap
registers the escaping ions during such a controlled reduction of
the trap rf amplitude, leading to an ion extraction mass spectrum
as shown in Fig.~\ref{fig:figure3}. Although this technique is
destructive and less accurate than the radial excitation, it has
several advantages. It is fast (an extraction takes a few seconds)
and it works also with uncooled ions. Moreover, it provides
qualitative information on the ion temperatures as the mass
spectra of warmer ion ensembles are broadened due to the ions'
wider energy distribution. This is shown in Fig.~\ref{fig:figure3}
for \baplus/\cyts~ensembles in the cooled (black) and the
noncooled case (gray). In the cooled case, both the peak of the
laser-cooled \baplus~ions and the peak of the \cyts~ions are
narrower. This is a qualitative proof of their sympathetic
cooling.

\begin{figure}
    \includegraphics[width=7.25cm]{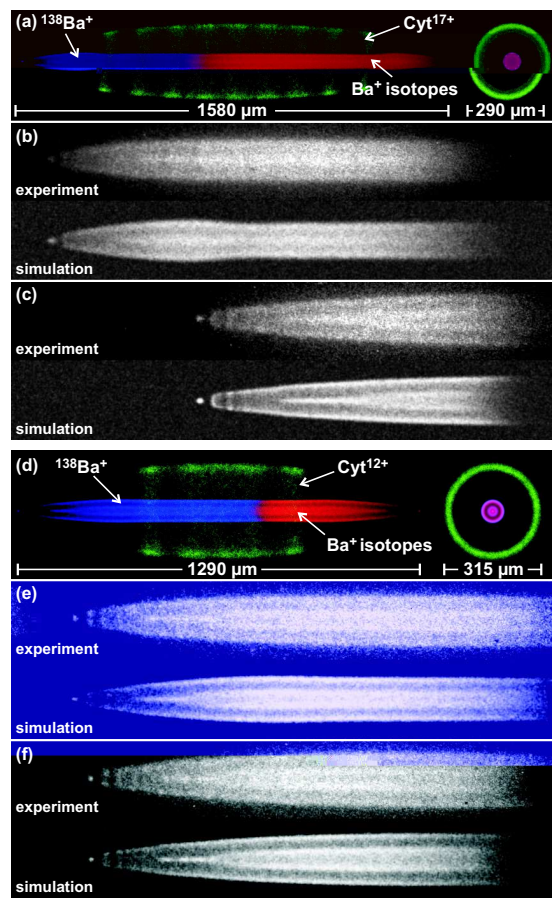}
    \caption{\label{fig:figure4} (color online) Experimental and
    simulated CCD images of a \baplus/\cyts~and a \baplus/
    \cytt~ion crystal. (a,~d) Simulated images in radial (left)
    and axial view (right), as they would appear if all ions would
    fluoresce. The cooling lasers propagate to the left, separating
    the \barium~ions (blue) and the barium isotopes (red) axially
    due to the light pressure force. The cytochrome ions (green)
    form a sheath around the barium ion subensemble. (b,~e)
    Experimental image of the barium-cytochrome ion crystal and its
    simulation, showing only the fluorescing \barium~ions.
    (c,~f) Crystals after the removal of the cytochrome
    ions. Here, the barium ion ensemble is no longer deformed nor
    heated by the molecular ions.}
\end{figure}

\baplus/\cytn~ crystals were routinely produced. Several
\baplus/\cytt~and \baplus/\cyts~ion crystals have been analyzed
quantitatively by the use of MD simulations in order to obtain the
numbers of the trapped ions and their translational temperatures
$T_{\textrm{sec},i}=\frac{2}{3}\langle E_{i}\rangle/k_{B}$, where
$\langle E_{i}\rangle$ is the time- and subensemble-averaged
secular kinetic energy per ion of the species $i$. Our MD method
is described in \cite{Ostendorf2006} and \cite{Zhang2007}. Here,
we do not consider micromotion in the simulations explicitly since
it does not affect the CCD images \cite{footnote2a}.
Micromotion-related effects, such as rf heating, are taken into
account phenomenologically.

In a quadrupole trap ions of different charge-to-mass ratios
arrange radially separated according to their effective radial
trap potentials $\Phi(r)\propto \frac{Q}{m}r^{2}$ \cite{footnote1}
and the interspecies repulsion $\sim Q_{1}Q_{2}$. In the limit of
cylindrical symmetry, this radial separation scales as
$r_{1}/r_{2}=(Q_{2}m_{1}/Q_{1}m_{2})^{1/2}$ with the outer radius
$r_{1}$ of the lower mass-to-charge ratio $(m_{1}/Q_{1})$
subensemble and the inner radius $r_{2}$ of the higher
mass-to-charge ratio $(m_{2}/Q_{2})$ subensemble
\cite{Hornekaer2000}. As shown in Figs.~\ref{fig:figure4}~(a)
and~\ref{fig:figure4}~(d), the multiply charged cytochrome ions
form a sheath around the barium ions which radially squeezes and
axially prolongs the barium ion subensemble. In the experiment two
CCD images of the ion crystals are recorded: one with the
molecular ions present and one after the molecular ions have been
removed. In the case presented in Figs.~\ref{fig:figure4}~(a)$-$
\ref{fig:figure4}~(c) the \cyts~ions prolong the barium ion
subensemble axially by 240~$\mu$m [Fig.~\ref{fig:figure4} (b) and
~\ref{fig:figure4}~(c)], which can be explained by the presence of
53 \cyts~ions, according to MD simulations. The barium ensemble
consists of 158 laser-cooled \barium~ions and 230 barium isotopes
other than $^{138}$Ba. Due to nonoptimal trap parameters during
the experiment of Fig.~\ref{fig:figure4}~(b), the observed radial
squeezing of the barium ion subensemble is less pronounced than in
the simulated image.

For the temperature determination a specific heating rate
$h_{i}=d\langle E_{i}\rangle/dt$ for each ion species $i$ is
chosen that phenomenologically describes heating effects such as
collisions with background gas particles, rf heating, or electric
field noise \cite{Zhang2007}. For the laser-cooled \barium~ions a
cooling rate is employed corresponding to the laser cooling force
$F_{D}=-\beta v_{z}$, where $\beta$ is the damping coefficient and
$v_{z}$ the $z$ component of the ion velocity with the cooling
lasers propagating in the positive $z$ direction
\cite{Ostendorf2006}. The other species are cooled sympathetically
through the interparticle Coulomb interaction, resulting in a
steady state in which every ion species $i$ has its specific
translational temperature $T_{\textrm{sec},i}$. For the pure
\baplus-ion crystal shown in Fig.~\ref{fig:figure4}~(c) a heating
rate $h_{\textrm{Ba}^{+}}=\frac{3}{2}k_{B}(3.7$~K/s)
\cite{footnote3} needs to be applied to reproduce the experimental
CCD image concerning shape and internal structure of the
\barium~ion subensemble. Here, a realistic damping coefficient
$\beta=1.7\times10^{-22}$~kg/s was assumed \cite{Raab2000}, which
yields the temperatures
$T_{\textrm{sec},^{138}\textrm{Ba}^{+}}=19$~mK and
$T_{\textrm{sec},\textrm{Ba}^{+}\textrm{isotopes}}=23$~mK. As all
conditions remained constant, the same damping coefficient $\beta$
and the same heating rate $h_{\textrm{Ba}^{+}}$ are applied to
reproduce the CCD image of the corresponding "parent"
\baplus/\cyts~ion crystal [Fig.~\ref{fig:figure4}~(b)]. Here, a
heating rate $h_{\textrm{Cyt}^{17+}}=\frac{3}{2}k_{B}(71.6$~K/s)
must be present in order to reproduce the observed warmer barium
ion subensemble whose temperatures are
$T_{\textrm{sec},^{138}\textrm{Ba}^{+}}=66$~mK and
$T_{\textrm{sec},\textrm{Ba}^{+}\textrm{isotopes}}=109$~mK. With
this value, the cytochrome temperature is
$T_{\textrm{sec},\textrm{Cyt}^{17+}}=704$~mK. As the laser cooling
efficiency varies depending on numerous factors, we consider a
range of $\beta$ values, as observed by \cite{Bushev2004}. To
obtain a lower (upper) limit for the heating rates and
temperatures we simulate the cases
$\beta=1.0(2.0)\times10^{-22}$~kg/s. For the ion crystal shown in
Fig.~\ref{fig:figure4}~(b) with
$T_{\textrm{sec},^{138}\textrm{Ba}^{+}}=66$~mK, this yields the
following values:
$h_{\textrm{Ba}^{+}}=\frac{3}{2}k_{B}(3.7^{+0.6}_{-1.6}$~K/s),
$T_{\textrm{sec},\textrm{Ba}^{+}\textrm{isotopes}}=109^{+6}_{-17}$~mK,
$h_{\textrm{Cyt}^{17+}}=\frac{3}{2}k_{B}(71.6^{+11.9}_{-29.2}$~K/s),
and $T_{\textrm{sec},\textrm{Cyt}^{17+}}=704^{+26}_{-24}$~mK.

The temperature of the molecular ions mainly depends on the laser
cooling power, the pressure in the vacuum chamber, and the ratio
of the numbers of laser-cooled and sympathetically cooled ions and
also their mass-to-charge ratio compared to that of the
laser-cooled \barium~ions. The closer the ratios are, the lower is
the radial separation of the two species in the trap and the
higher is the sympathetic cooling efficiency. In the cases shown
in Fig.~\ref{fig:figure4}, the radial gap $(r_{2}-r_{1})$ between
the \cyts~and the \barium~ion ensembles is 64~$\mu$m and in the
\cytt~case 87~$\mu$m. This contributes to the fact that the
\cytt~ions could only be cooled to 1.2 K, although the barium ions
at 26 mK were much colder than in the \cyts~case \cite{footnote4}.
As a consequence of the different cytochrome temperatures, the
\cytt~subensemble [Fig.~\ref{fig:figure4}~(d)] appears more
blurred in the simulated images than the colder \cyts~subensemble
[Fig.~\ref{fig:figure4}~(a)].

In conclusion, we have shown that heavy and thus complex
particles, appropriately charged, can be translationally cooled in
the gas phase to temperatures significantly below those reachable
with other techniques. The lowest temperature observed was below
0.75~K. The temperature is limited by heating experienced by the
molecular ions. Its origin is uncertain so far, but may be due to
electrical noise. Combining the present result with earlier work
(sympathetic cooling of masses up to 410 amu)
\cite{Roth2005,Ostendorf2006} we reasonably expect that ions of
any mass between 32 and 13000 amu can be sympathetically cooled,
provided they have an adequately high charge. The particular
advantage of this cooling method is that the ions can be stored
for more than 20 min (easily extendable to hours using an improved
vacuum to avoid loss reactions of the \barium~ions) with a strong
spatial localization in a well-defined and nearly collisionless
environment. These are ideal conditions for, e. g., studies of
low-energy processes, like cold collisions or triplet-singlet
decay, or high-resolution spectroscopy for structural or
conformational characterizations of biomolecules. For many
experiments cooling of the internal degrees of freedom of the
molecules would be favorable and could be implemented by radiative
cooling in a cryogenic environment.

We thank T. Schneider for helpful discussions. D.O. acknowledges
support from the Studienstiftung des deutschen Volkes and C.B.Z.
from the Deutscher Akademischer Austauschdienst (DAAD).


\begin{thebibliography}{19}
\bibitem{footnote2} The relationship
$U^{\textrm{ex}}_{\textrm{rf}}\propto \sqrt{m/Q}$ was derived from
an analysis of the stability diagram \cite{Drewsen2000} with
regard to the decrease of $q$ below stability when reducing
$U_{\textrm{rf}}$. It holds when the edge of the stability range
is approximated by $q=-\frac{1}{2}a^{2}$ (adequate for the range
of $q$ considered here).
\bibitem{footnote2a} Here, the micromotion energy
\cite{Hornekaer2000} is 25~meV for \cytt~and 38~meV for \cyts.
However, for spectroscopy experiments with laser beams parallel to
the trap axis this purely radial micromotion should not be
relevant.
\bibitem{footnote3} Due to their low mass difference, the heating
rates for the \barium and the barium isotope ions are assumed
equal.
\end{thebibliography}
\end{document}